\PassOptionsToPackage{table}{xcolor}
\documentclass[sigconf]{acmart} 

\usepackage{multirow}
\usepackage{amsmath}
\usepackage{mathtools}
\usepackage{graphicx}
\usepackage{listings}
\usepackage{verbatim}
\usepackage[table]{xcolor}
\newcommand{\graydashline}{%
  \noalign{\vskip 2pt}%
  \noalign{\color{gray!40}\leaders\hbox to 3pt{\hss.\hss}\hfill\kern0pt}%
  \noalign{\vskip 2pt}%
}
\usepackage{colortbl}
\definecolor{cB}{HTML}{B0B0B0} 
\definecolor{cK}{HTML}{948979} 
\definecolor{cH}{HTML}{AAABF7} 
\newcommand{\CB}{{\setlength{\fboxsep}{1.5pt}\colorbox{cB}{\textcolor{white}{\sffamily\small\bfseries B}}}}
\newcommand{\CK}{{\setlength{\fboxsep}{1.5pt}\colorbox{cK}{\textcolor{white}{\sffamily\small\bfseries K}}}}
\newcommand{\CH}{{\setlength{\fboxsep}{1.5pt}\colorbox{cH}{\textcolor{white}{\sffamily\small\bfseries H}}}}

\definecolor{cCog}{HTML}{534AB7}
\usepackage[utf8]{inputenc}
\usepackage{appendix}

\lstdefinestyle{promptB}{
  backgroundcolor=\color{cB!5}, frame=single, framerule=0.3mm,
  rulecolor=\color{cB!50}, xleftmargin=3pt, xrightmargin=3pt,
  aboveskip=2pt, belowskip=6pt
}
\lstdefinestyle{promptK}{
  backgroundcolor=\color{cK!5}, frame=single, framerule=0.3mm,
  rulecolor=\color{cK!50}, xleftmargin=3pt, xrightmargin=3pt,
  aboveskip=2pt, belowskip=6pt
}
\lstdefinestyle{promptH}{
  backgroundcolor=\color{cH!5}, frame=single, framerule=0.3mm,
  rulecolor=\color{cH!50}, xleftmargin=3pt, xrightmargin=3pt,
  aboveskip=2pt, belowskip=6pt
}
\AtBeginDocument{%
  }

\copyrightyear{2026}
\acmYear{2026}
\setcopyright{cc}
\setcctype{by-nc-nd}
\acmConference[UIST '26]{The 39th Annual ACM Symposium on User Interface Software and Technology}{November 02--05, 2026}{Detroit, MI, USA}
\acmBooktitle{The 39th Annual ACM Symposium on User Interface Software and Technology (UIST '26), November 02--05, 2026, Detroit, MI, USA}
\acmDOI{10.1145/3830398.3830724}
\acmISBN{979-8-4007-2856-3/2026/11}

\begin{document}

\title[CogChat: Knowledge Graph-Augmented Conversational AI \newline with Heterogeneous Graph Transformer for Cognitive Grounding in Design Generation]{CogChat: Knowledge Graph-Augmented Conversational AI with Heterogeneous Graph Transformer for Cognitive Grounding in Design Generation}

\author{Jiin Choi}
\affiliation{
  \department{Design AI Lab}
  \institution{Hanyang University}
  \city{Seoul}
  \country{Republic of Korea}
}

\affiliation{
  \department{Human-Centered AI Design Institute}
  \institution{Hanyang University}
  \city{Seoul}
  \country{Republic of Korea}
}
\email{jiin4900@gmail.com}
\orcid{0009-0003-2513-6895}

\author{Kyung Hoon Hyun}
\authornote{Corresponding author.}
\affiliation{
  \department{Design AI Lab}
  \institution{Hanyang University}
  \city{Seoul}
  \country{Republic of Korea}
}
\affiliation{
  \department{Human-Centered AI Design Institute}
  \institution{Hanyang University}
  \city{Seoul}
  \country{Republic of Korea}}
\email{hoonhello@gmail.com}
\orcid{0000-0001-6379-9700}

\begin{abstract}
LLM-based chat systems have become valuable tools for design practice, enabling rapid ideation and flexible task support. Yet these systems process designer utterances as generic sequences, maintaining context through recency rather than through any model of how the speaker organizes knowledge. In design conversation, this gap compounds as relational context decays between turns, identical words go unresolved across designers, and the conversation loops or restarts rather than deepens. We present \textsc{\textbf{\textcolor{cCog}{CogChat}}}, a real-time chat framework that grounds conversational AI in a personal heterogeneous knowledge graph constructed from each designer's input. The system extracts typed entities and relations into a heterogeneous graph, then applies a HGT (Heterogeneous Graph Transformer) to select structurally relevant nodes for response generation and to generate both intentional and exploratory probing questions. Technical evaluation shows that HGT-based entity selection outperforms both ungrounded LLM interaction and naïve KG augmentation, which introduces noise that degrades response quality. A within-subjects study with nine professional designers indicates that grounding conversation in a relationally structured, designer-specific semantic context improves context retention, personalized intent interpretation, and conversational depth while reducing cognitive load. These findings suggest that structuring a designer's expressed concepts and relations as a dynamic knowledge graph can preserve relational context that fades across turns, pointing toward a graph-grounded approach to long-term context management in LLM-based interaction.
\end{abstract}

\ccsdesc[500]{Human-centered computing~Interactive systems and tools}
\ccsdesc[500]{Computing methodologies~Information extraction}
\ccsdesc[500]{Computing methodologies~Ontology engineering}
\ccsdesc[300]{Computing methodologies~Machine learning}
\keywords{Knowledge Graph, Heterogeneous Graph Transformer, Design Cognition, Conversational AI}
\begin{teaserfigure}
  \includegraphics[width=\textwidth]{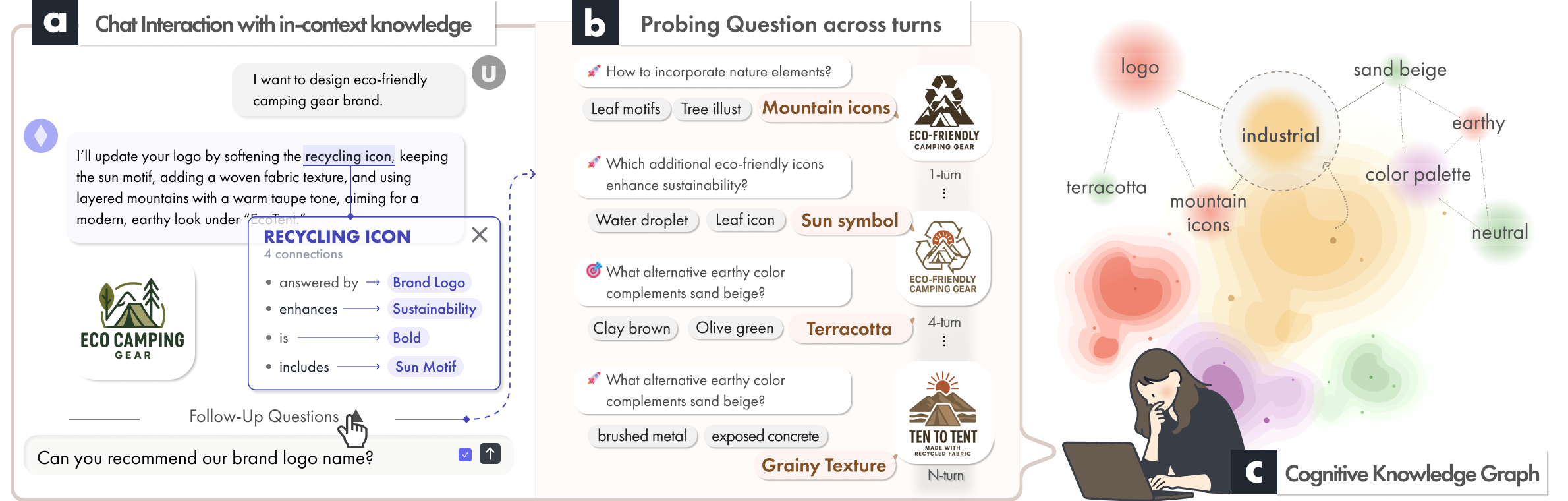}
\caption{Overview of \textsc{\textbf{\textcolor{cCog}{CogChat}}}. (a) Chat interaction with in-context knowledge lookup. (b) Probing questions with suggested answers and progressively refined outputs across turns. (c) Cognitive knowledge graph constructed from designer input.}
\Description{Three-panel system overview of CogChat. (a) Chat with in-context knowledge popup showing the term "recycling icon" and its four graph relations: answered_by Brand Logo, enhances Sustainability, is Bold, and includes Sun Motif. (b) Probing questions across turns with suggested answers and progressively refined logo outputs. (c) Cognitive knowledge graph with interconnected design concepts such as earthy, terracotta, mountain icons, and color palette.}
  \label{fig:teaser}
\end{teaserfigure}


\maketitle
\section{Introduction}

Recent conversational AI systems have extended language model capabilities through retrieval-augmented generation~\cite{lewis2020retrieval,cheng2025ragtrace,mei2025interquest}, agent-based reasoning~\cite{yao2022react,peng2025morae,gunturu2025mapstory}, and memory management~\cite{packer2023memgpt,zhong2024memorybank,pu2025promemassist}. These approaches improve factual grounding and task decomposition, but share a structural limitation: the user remains a generic interlocutor. Context is maintained through recency, not through any model of how the user organizes knowledge~\cite{laban2025llms,dongre2025drift}.

In design practice, this limitation produces three compounding failures. First, \textit{information loss}: relational context between turns decays as conversation advances. The system forgets that ``warm'' referred to tactile grain, not color temperature---the kind of situated meaning central to design reasoning~\cite{schon1983reflective}. Second, \textit{personalization failure}: identical words carry different meanings across designers---``heavy'' as visual density versus information overload---but the system processes only generic semantics~\cite{cross2011understanding}. Third, \textit{inefficiency}: without retained context or resolved meaning, designers re-explain established ground, and the conversation loops rather than deepens~\cite{goldschmidt2014linkography}. These are not independent problems but symptoms of a single absence: no persistent, structured representation of the designer's cognitive space. Design cognition is heterogeneous---perceptual, semantic, and functional elements coexist within a single judgment~\cite{lawson2006designers}---and flattening this into token sequences discards the relational structure that makes design reasoning coherent.

We present \textsc{\textbf{\textcolor{cCog}{CogChat}}}, a chat framework that grounds LLM interaction in a personal heterogeneous knowledge graph~(KG) constructed in real time from each designer's input (Figure~\ref{fig:teaser}). Rather than modeling a designer's internal reasoning or latent cognition, \textsc{\textbf{\textcolor{cCog}{CogChat}}} maintains a designer-specific relational context across turns, contributing graph-grounded conversational personalization through structured relational memory. The framework addresses four requirements mapped to these three failures: (1)~\textit{Structure} preserves explicit relations among utterances and concepts, countering information loss; (2)~\textit{Represent} embeds heterogeneous elements in a type-preserving space via Heterogeneous Graph Transformer (HGT)~\cite{hu2020heterogeneous}, enabling personalized interpretation; (3)~\textit{Act} reads the graph's structural state to generate grounded responses and two types of probing questions---\textit{intentional questions} targeting knowledge gaps and \textit{exploratory questions }expanding the conceptual frontier; and (4)~\textit{Evolve} updates the graph incrementally, sustaining all three across turns (\texttt{See Details in Supplementary Video}).

We evaluate three conditions---\textbf{\textit{Baseline}} (LLM alone), \textbf{\textit{KG-only}}, and \textbf{\textit{KG+HGT}}--- to examine whether grounding conversation in a designer-specific relational context improves both system performance and user experience. Technical benchmarks show that HGT-based entity selection outperforms both ungrounded interaction and na\"{i}ve full-KG augmentation, which degrades performance by injecting noise. A within-subjects study with nine professional designers then asks whether this technical advantage reshapes actual design conversation, addressing three questions: whether graph-based grounding improves cross-turn context retention (\textbf{RQ1}); whether a heterogeneous KG captures designer-specific language beyond generic semantics (\textbf{RQ2}); and whether structurally grounded interaction lowers the cognitive effort of communicating intent and deepens engagement (\textbf{RQ3}).

The key contributions of our research include:
\begin{enumerate}
    \item We propose a real-time chat architecture that grounds LLM interaction in a personal heterogeneous knowledge graph constructed from each designer's input, structuring the relations a designer explicitly expresses rather than relying on recency-based retrieval.
    \item We introduce a probing mechanism that partitions the graph's embedding space into intentional and exploratory zones, generating structurally grounded questions.
    \item We conducted technical benchmarks and a within-subjects designer study showing that HGT-based selective grounding outperforms both ungrounded and na\"{i}vely grounded alternatives in context retention, personalization, and conversational depth.
\end{enumerate}

\section{Related Work}

\subsection{LLM-based Conversational Systems}

Large language models have advanced conversational AI through coherent multi-turn dialogue, instruction following, and task decomposition~\cite{achiam2023gpt,touvron2023llama}. Retrieval-augmented generation (RAG) grounds responses in external documents to reduce hallucination~\cite{lewis2020retrieval}, while agent-based architectures introduce multi-step reasoning and tool use~\cite{yao2022react}. Memory-augmented approaches such as MemGPT~\cite{packer2023memgpt} and MemoryBank~\cite{zhong2024memorybank} extend conversational continuity through hierarchical memory management, and context window expansion has further enabled longer interaction histories.

Despite these advances, a shared limitation persists: the user is treated as a generic interlocutor. Context is maintained through recency or retrieval proximity---not through any representation of how the speaker organizes knowledge. Recent work confirms that this limitation is not merely theoretical: LLMs exhibit significant performance degradation in multi-turn settings, making early assumptions and failing to integrate new information as conversation progresses~\cite{laban2025llms}. These systems track what was said recently, but not who is saying it, what conceptual categories they use, or how they relate ideas.

\begin{figure*}[!htbp]
  \centering
  \includegraphics[width=\textwidth]{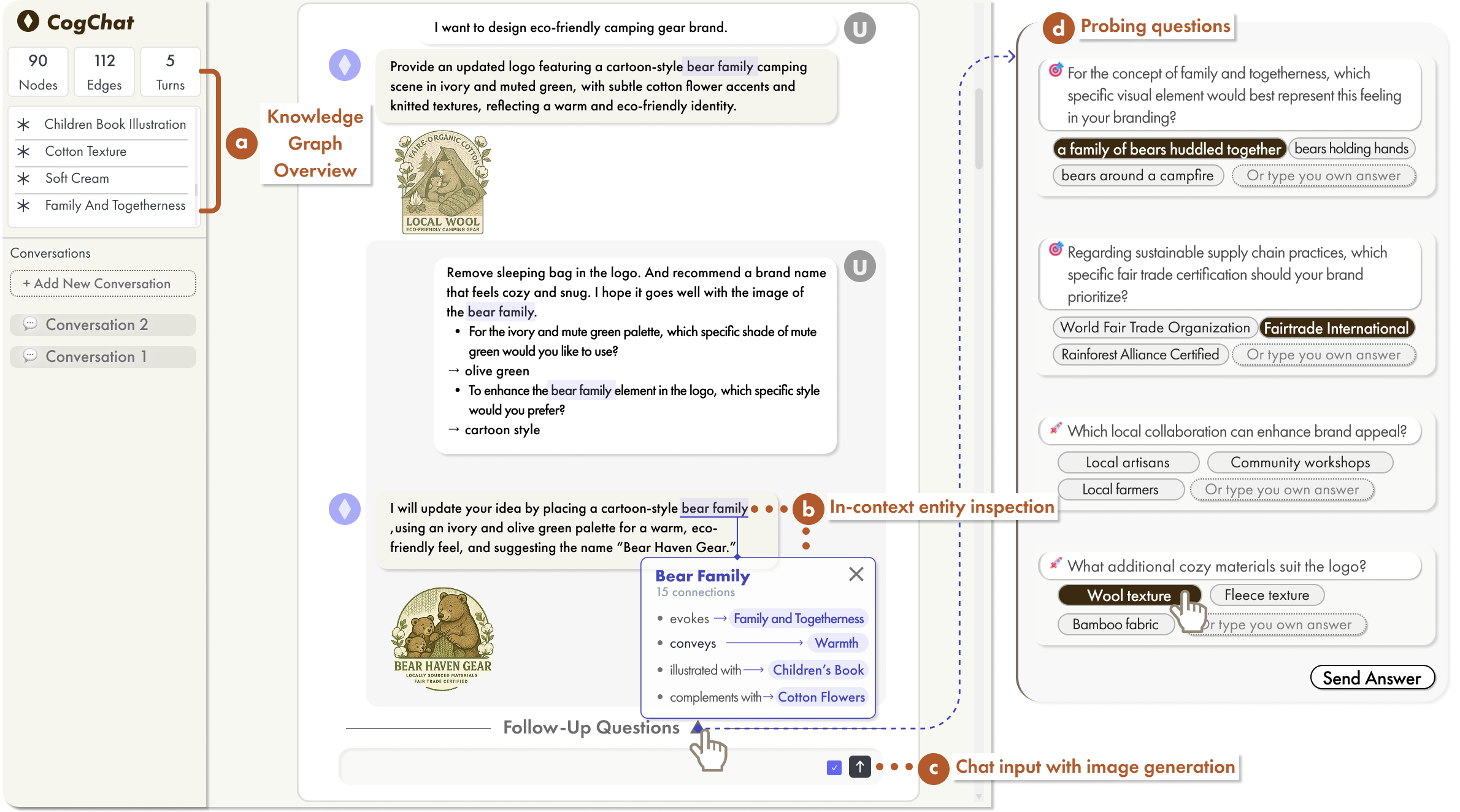}
  \caption{\textsc{\textbf{\textcolor{cCog}{CogChat}}} interface. (a) Node and edge counts with extracted entity chips. (b) Double-clicking a term reveals its graph relations. (c) Text input with image generation toggle. (d) Intentional and exploratory questions with suggested answers.}
  \Description{CogChat interface composed of four panels. (a) A sidebar displaying node and edge counts alongside extracted entity chips representing concepts from the designer's input. (b) A graph relation view triggered by double-clicking a term, showing connected nodes and typed edges. (c) A text input field with an image generation toggle for multimodal interaction. (d) A probing question panel presenting intentional questions targeting knowledge gaps and exploratory questions expanding the conceptual frontier, each accompanied by suggested answers.}
  \label{fig:2}
\end{figure*}

\subsection{User Modeling in Design Tools and Design Cognition}

User modeling in design tools has largely relied on behavioral signals. Recommendation systems in platforms such as Pinterest and Canva adapt to implicit feedback---clicks, saves, dwell time---to surface stylistically relevant content~\cite{zhai2017visual}. In conversational systems, personalization has been approached through user profile embeddings~\cite{zheng2020pre} and preference-conditioned generation~\cite{salemi2024lamp}. These models operate on aggregate behavioral patterns rather than individual cognitive structure. Recent work suggests that question-asking itself can scaffold cognitive engagement in AI-assisted workflows, indicating that probing questions are effective only when grounded in the individual's knowledge structure~\cite{gmeiner2025exploring,lim2024co}. Gmeiner et al.~\cite{gmeiner2025exploring} found that reflective questions improved intent formulation and problem exploration in design, while Lim et al.~\cite{lim2024co} showed that LLM-generated questions helped authors articulate intent but failed to capture individual voice without structured involvement.

Design cognition research has established that designers differ systematically in how they categorize visual form, weight semantic relationships, and apply domain heuristics~\cite{cross2011understanding,lawson2006designers}. Sch\"{o}n's reflection-in-action~\cite{schon1983reflective} and Goldschmidt's linkography~\cite{goldschmidt2014linkography} demonstrate that design reasoning is structured, relational, and individually distinct. Yet these insights have not been operationalized in interactive systems. The gap is consistent: existing tools model what designers \textit{do}, not how they \textit{think}. No current approach structures a designer's individual knowledge space and grounds real-time conversation in that representation.

\subsection{Knowledge Graphs and Heterogeneous Graph Representation}

Knowledge graphs have proven effective for grounding language model outputs in structured relational knowledge. KGQA systems leverage entity-relation triples to improve answer precision~\cite{saxena2021question}, and commonsense KGs such as ConceptNet~\cite{speer2017conceptnet} and Wikidata~\cite{vrandevcic2014wikidata} provide relational scaffolding that LLMs can query at inference time. Personal knowledge graph research has explored structuring individual data---emails, documents, calendar events---into queryable graphs~\cite{balog2019personal,semnani2023wikichat}. However, these graphs capture what a person has \textit{done}, not how they \textit{structure knowledge}.

Graph-based learning has evolved to handle increasingly complex relational structures. GCN~\cite{kipf2016semi} established neighborhood aggregation; GAT~\cite{velivckovic2017graph} introduced attention-weighted neighbor selection; R-GCN~\cite{schlichtkrull2018modeling} added relation-specific weight matrices. These models, however, treat node and edge types uniformly, limiting their capacity to distinguish semantically distinct entities. HGT~\cite{hu2020heterogeneous} addresses this through type-specific linear projections and relation-specific attention, computing separate Query/Key/Value transformations per node type. Design cognition---where perceptual segments, semantic attributes, and functional properties coexist within a single judgment---forms precisely the kind of heterogeneous, densely connected graph that HGT is designed for. No existing work constructs such a graph from a designer's cognitive patterns and integrates it into a live conversational interface.

\begin{figure*}[!htbp]
  \centering
  \includegraphics[width=\textwidth]{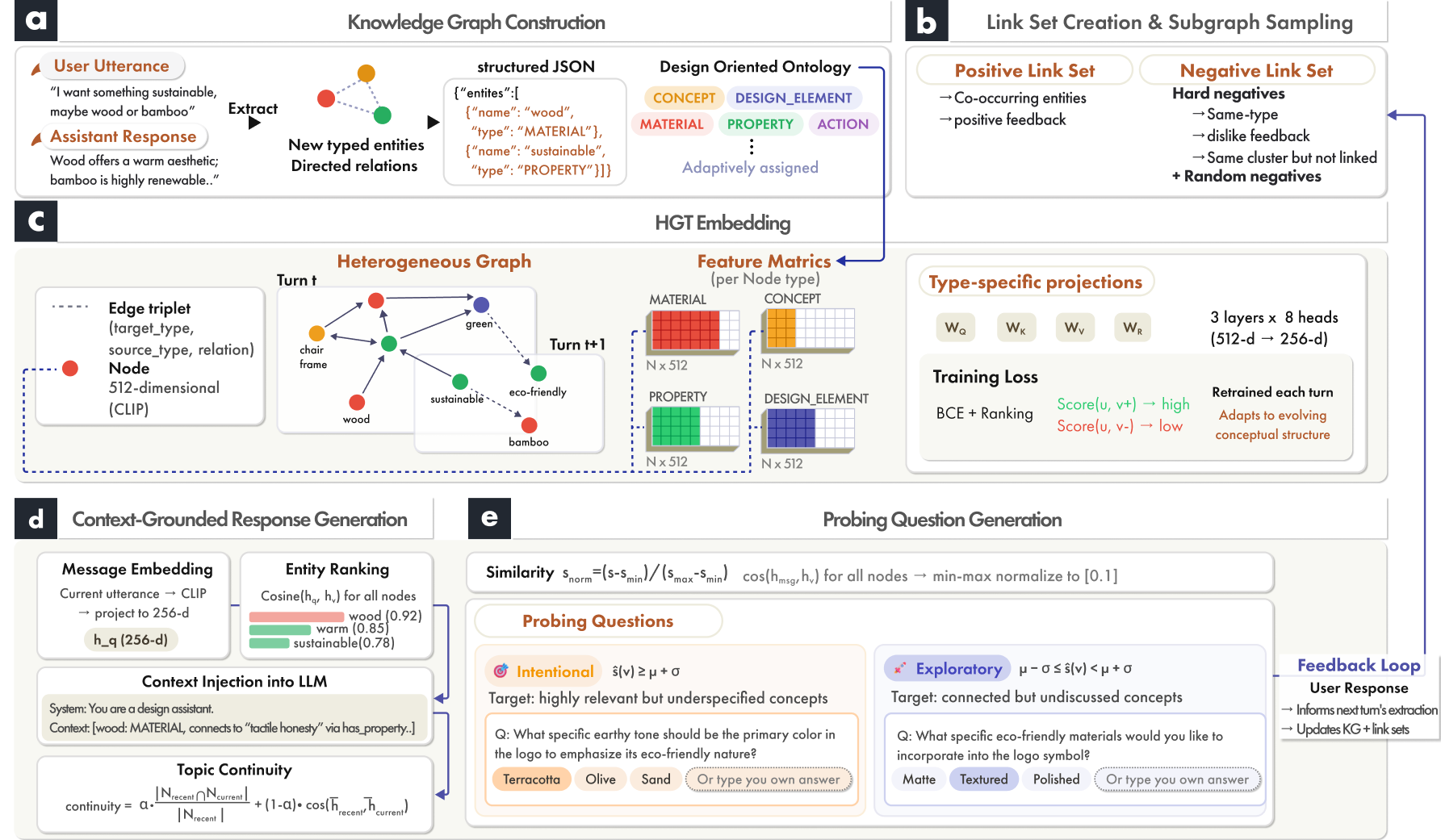}
  \caption{System pipeline. (a) Entity and relation extraction. (b) Positive/negative link set creation with subgraph sampling. (c) HGT embedding with type-specific projections. (d) Embedding-based entity ranking and context injection into LLM. (e) Partitioning for intentional and exploratory probing questions, with responses feeding back into extraction.}
  \Description{Five-stage system pipeline diagram. (a) Entity and relation extraction from designer utterances producing structured JSON. (b) Positive and negative link set creation with subgraph sampling around candidate nodes. (c) HGT embedding with type-specific linear projections applied per node type. (d) Embedding-based entity ranking and injection of selected context into the LLM for response generation. (e) Partitioning of graph state into intentional and exploratory probing questions, with designer responses looping back into extraction.}
  \label{fig:3}
\end{figure*}

\section{\textsc{\textbf{\textcolor{cCog}{CogChat}}} System}

\subsection{Interface Walkthrough}

The interface integrates conversation, knowledge inspection, and probing into a single workspace (Figure~\ref{fig:2}), comprising four key components: a knowledge graph overview (Figure~\ref{fig:2}a), in-context entity inspection (Figure~\ref{fig:2}b), chat input with image generation (Figure~\ref{fig:2}c), and probing questions with suggested answers (Figure~\ref{fig:2}d).

\subsubsection{Chat and Image Generation}
Designers converse with an LLM assistant that is continuously grounded in their own dynamic knowledge graph (Figure~\ref{fig:2}c). Each message builds on the accumulated relational context from prior turns---entities extracted across the conversation are structured into a heterogeneous KG and embedded via HGT, so the system's responses reflect not just the current utterance but the designer's evolving conceptual structure. A checkbox next to the input field toggles image generation---when enabled, the system produces a visual informed by the HGT-embedded conversational context rather than the surface text alone. Multiple sessions can be opened from the left sidebar, each maintaining an independent knowledge graph that persists across turns. The left panel displays accumulated node and edge counts alongside extracted entity chips (Figure~\ref{fig:2}a).

\subsubsection{Probing Questions}
To reduce ambiguity and acquire richer context for knowledge graph construction, the system presents probing questions derived from the graph's current structural state (Figure~\ref{fig:2}d).  \includegraphics[height=1em]{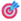}\textit{Intentional questions} target gaps or ambiguities the graph has identified---if the designer said ``earthy tones'' without specifying further, the system asks which specific tone. \includegraphics[height=1em]{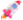}\textit{Exploratory questions} surface connections the designer has not yet discussed but that the graph topology suggests may be relevant. Each question includes 2--3 suggested answers; the designer's response feeds directly into the next turn's extraction, deepening the graph and tightening the system's model of their intent across subsequent turns.

\subsubsection{In-Context Knowledge Lookup}
Designers can double-click any word in the chat thread to surface its structural context within the personal knowledge graph (Figure~\ref{fig:2}b). Without leaving the conversation, a summary popup shows the entity's connection count and direct relations---for instance, selecting ``bear family'' reveals 15 connections including \textit{evokes} $\rightarrow$ Family and Togetherness, \textit{conveys} $\rightarrow$ Warmth, \textit{illustrated with} $\rightarrow$ Children's Book, and \textit{complements with} $\rightarrow$ Cotton Flowers. This lets designers check whether the system's accumulated understanding of a term aligns with their own, and catch misinterpretations before they propagate through future turns.

\subsection{Technical Pipeline}

The pipeline executes five stages at every conversational turn: knowledge graph construction, link set creation, subgraph sampling and HGT embedding, context-grounded response generation, and probing question generation (Figure~\ref{fig:3}). Each stage feeds the next, so the designer's accumulated conceptual structure is continuously updated and reflected in real time.

\subsubsection{Knowledge Graph Construction}
At each turn, the system extracts typed entities and directed relations from both the user utterance and the assistant response based on GraphRAG~\cite{edge2024local}, producing structured JSON (Figure~\ref{fig:3}a). Each entity is assigned a type guided by a design-oriented ontology---\texttt{CONCEPT}, \texttt{DESIGN\_ELEMENT}, \texttt{MATERIAL}, \texttt{PROPERTY}, and \texttt{ACTION}---with the model adapting beyond these categories as conversational content demands. To mitigate hallucination propagation, the graph is refined at each turn, allowing designers to review and correct extracted relations directly.

The extracted output is instantiated as a heterogeneous graph. Each node type maintains a separate feature matrix initialized with 512-dimensional CLIP embeddings (ViT-B/32), and edges are indexed by (\texttt{target\_type}, \texttt{source\_type}, \texttt{relation}) triplets. The graph grows incrementally: at each turn, new entities merge into the existing structure through string matching and embedding similarity. Cross-turn relations connect question entities to answer entities via \texttt{answered\_by} edges, explicitly preserving the relational link between what the designer asked and how the system responded. Over successive turns, the graph accumulates both user-side nodes (utterances, preferences, intents) and value-side nodes (design elements, materials, attributes, style properties) into a unified heterogeneous structure. 

\subsubsection{Link Set Creation and Subgraph Sampling}
To train the HGT, the system constructs positive and negative link sets from the accumulated graph~\cite{ying2018graph} (Figure~\ref{fig:3}b).

Positive link sets are drawn from entity pairs that co-occur within the same utterance or that the designer explicitly reinforces through positive feedback. For example, if an utterance produces the entities \texttt{CHAIR}, \texttt{ARMREST}, and \texttt{RED}, the pairs (\texttt{CHAIR}, \textit{has\_part}, \texttt{ARMREST}) and (\texttt{ARMREST}, \textit{has\_property}, \texttt{RED}) form positive links.

Negative sampling generates contrastive pairs through two strategies. \textit{Hard negatives} are sampled from structurally proximate but unconnected nodes: same-type unchosen entities (e.g., other \texttt{MATERIAL} nodes not selected by the designer), entities exposed in conversation but rejected via dislike feedback, and entities within the same HDBSCAN cluster but not directly linked. \textit{Random negatives} are sampled uniformly from nodes beyond the two-hop neighborhood. Edge weights adjust with feedback---positive feedback increases connection strength, negative feedback decreases it.

The system then performs neighborhood sampling around entities mentioned in the current utterance, extracting 1-hop and 2-hop subgraphs from both sides of the knowledge graph~\cite{yan2024knownet}. Sampling follows a random walk and layer-wise strategy: at each hop, neighbors are sampled proportionally to edge weight, ensuring that feedback-reinforced connections are more likely to be included. This produces a compact subgraph preserving local relational structure while keeping the input tractable for real-time embedding.

\subsubsection{HGT Embedding}
The sampled subgraph is embedded using a Heterogeneous Graph Transformer (HGT)~\cite{hu2020heterogeneous} (Figure~\ref{fig:3}c). HGT computes type-specific Query, Key, and Value projections for each node type and relation-specific attention weights for each edge type. For node $v$ of type $\tau(v)$ connected to neighbor $u$ of type $\tau(u)$ through relation $\phi(e)$:

\begin{equation}
  \text{Attention}(v, e, u) = \text{softmax}\!\left(\frac{(W^{Q}_{\tau(v)} h_v)^{\top}(W^{K}_{\tau(u)} h_u)}{\sqrt{d_k}} \cdot W^{R}_{\phi(e)}\right)
\end{equation}

where $W^Q$, $W^K$ are type-specific projection matrices and $W^R$ is a relation-specific weight matrix. The model uses 3 layers with 8 attention heads, producing 256-dimensional embeddings from 512-dimensional CLIP inputs.

The model trains at each turn on the current link sets using a combination of BCE loss and ranking loss. For a positive pair $(u, v^+)$ and negative pair $(u, v^-)$, the objective encourages $\text{Score}(u, v^+) \rightarrow \text{high}$ and $\text{Score}(u, v^-) \rightarrow \text{low}$, where Score is the dot product of HGT embeddings. Because the model retrains incrementally with updated link sets at every turn, the embedding space continuously adapts to the designer's evolving conceptual structure. For cold-start sparsity, \textsc{\textbf{\textcolor{cCog}{CogChat}}} bypasses HGT when the graph is small (< 5 nodes), injecting all nodes directly; HGT selection activates once enough structure emerges.

\subsubsection{Context-Grounded Response Generation}
Each node is ranked by HGT embedding similarity to the current message (Figure~\ref{fig:3}d). Given the message embedding $h_q$ (projected from CLIP to 256 dimensions) and each node's HGT embedding $h_v$:

\begin{equation}
  \text{rel}(v) = \cos(h_q, h_v) = \frac{h_q^{\top} h_v}{\|h_q\|\|h_v\|}
\end{equation}

The top-$k$ ($k\approx20$) entities are injected into the LLM prompt as structured context. HGT scores nodes by representation quality (embedding norm) and consistency with neighbors (inter-node similarity), so poorly extracted entities score low and fall out of the top-$k$---persisting in the graph but rarely injected. Each selected entity is accompanied by its type, key relations, and a natural-language summary of its subgraph neighborhood---providing the model with relational context rather than isolated facts. For instance, the prompt conveys not merely that the designer mentioned ``brutalist concrete'' but that this concept connects to ``tactile honesty'' and ``minimal ornamentation'' through specific graph paths established in earlier turns (Appendix \ref{app:2},~\ref{app:3}).

The system also tracks a continuity score that combines node set overlap with HGT embedding similarity:

\begin{equation}
  \text{continuity} = \alpha \cdot \frac{|N_{\text{recent}} \cap N_{\text{current}}|}{|N_{\text{recent}}|} + (1 - \alpha) \cdot \cos(\bar{h}_{\text{recent}},\, \bar{h}_{\text{current}})
\end{equation}

where $N_{\text{recent}}$ and $N_{\text{current}}$ are node sets from recent and current turns, $\bar{h}$ denotes averaged HGT embeddings, and $\alpha$ is set to weight embedding similarity higher than lexical overlap, since embeddings capture semantic continuity even when vocabulary shifts---a transition from ``concrete'' to ``plaster'' shares low node overlap but high embedding similarity, as both occupy structurally similar positions in the graph. When continuity drops below a threshold, the system appends a context warning to the prompt to maintain coherence while addressing the new direction.

\subsubsection{Probing Question Generation}
Prior work has shown that combining clarifying questions with divergent prompts deepens both intent articulation and creative exploration in design~\cite{yan2024knownet,lim2024identify}. Building on this, the system computes cosine similarity between the current message embedding and all node HGT embeddings, then applies min-max normalization (Figure~\ref{fig:3}e): $\hat{s}(v) = \frac{s(v) - s_{\min}}{s_{\max} - s_{\min}}$, where $s_{\min}$ and $s_{\max}$ are the minimum and maximum similarity values across all nodes in the current turn. Nodes are then partitioned into three zones based on the mean $\mu$ and standard deviation $\sigma$ of the normalized distribution:

\begin{equation}
  \begin{cases}
    \hat{s}(v) \geq \mu + \sigma & \rightarrow \quad \text{intentional} \\
    \mu - \sigma \leq \hat{s}(v) < \mu + \sigma & \rightarrow \quad \text{exploratory} \\
    \hat{s}(v) < \mu - \sigma & \rightarrow \quad \text{excluded}
  \end{cases}
\end{equation}

Nodes above $\mu + \sigma$ seed \textit{intentional questions}---these resolve ambiguity by targeting concepts that are highly relevant to the current utterance but underspecified in the graph. If the designer said ``earthy tones'' without elaborating, the system generates: \textit{``Which earthy base color---terracotta, olive, or sand?''} Nodes in the middle band seed \textit{exploratory questions}, probing regions that are structurally connected but not yet discussed, expanding the conversation into adjacent conceptual territory the designer may not have considered. Nodes below $\mu - \sigma$ are excluded as contextually irrelevant to the current turn. This distribution-based partitioning adapts dynamically to graph size and conversational stage---early in conversation when the graph is sparse, the thresholds are wider; as the graph densifies, the zones tighten around increasingly specific relevance boundaries.

Each question includes 2--3 suggested short answers. The designer's selection or free-text response feeds into the next turn's entity extraction: probing questions improve graph quality, which improves subsequent probing questions and response grounding (Appendix \ref{app:1}).

\subsection{Implementation Details}

The frontend is built with React and Vite; the backend runs on Python Flask with an NVIDIA RTX 4090 GPU on Linux. Graph extraction uses GPT-4o via GraphRAG, all conversational generation uses GPT-5.4 (\texttt{gpt-5.4-2026-03-05}), and image generation uses GPT Image 1.5 (\texttt{gpt-image-1.5-2025-12-16}). Graph operations and HGT embedding are handled by pyHGT~\cite{hu2020heterogeneous}, with CLIP providing initial 512-dimensional node features. Since HGT attends to relevant subgraph neighborhoods rather than the full graph, latency remains stable as the graph grows; average turn latency is approximately 2.3 seconds.

For evaluation, we deploy three system conditions sharing the same interface but differing in grounding depth:

\begin{itemize}
  \item \textbf{\textit{Baseline}} (\CB~LLM-only): receives the full conversation history within the model's context window and generates responses and follow-up questions from this unstructured dialogue record, with no knowledge graph extraction.
  \item \textbf{\textit{KG-only}} (\CK~LLM+KG): extracts entities into a knowledge graph and uses graph structure with keyword matching to generate structurally informed questions, but without embedding-based selection.
  \item \textbf{\textit{KG+HGT}} (\CH~LLM+KG+HGT): runs the full pipeline---graph construction, HGT embedding, selective entity injection, and distribution-based probing---producing responses and questions grounded in the designer's personal cognitive structure.
\end{itemize}

\section{Technical Evaluation}
\label{sec:4}
Before examining effects in design practice, we validate the core mechanism---whether HGT-based selection improves response quality on four benchmarks. The first two assess ambiguity and preference resolution: ASQA~\cite{stelmakh2022asqa} tests ambiguous questions admitting multiple valid interpretations, mirroring polysemous design utterances; RewardBench~\cite{lambert2025rewardbench} tests preference alignment under relational complexity. The other two assess long-term conversational memory: LoCoMo~\cite{maharana2024lococmo} evaluates recall across very long, multi-session dialogues; LongMemEval~\cite{wu2024longmemeval} tests temporal reasoning and information integration across sessions. For all four, KGs were constructed per-query from questions and reference passages using the same extraction pipeline.

\paragraph{Setup.} We compare three conditions: \textit{\textbf{Baseline}} (\CB~LLM-only with no KG context), \textit{\textbf{KG-only}} (\CK~LLM+KG, all extracted entities and their 1-hop and selective 2-hop neighbors provided as context), and \textit{\textbf{KG+HGT}} (\CH~LLM+KG+HGT, HGT-based scoring selects only top-$k$ ($k\approx20$) entities as context). All conditions use GPT-5.4 as the base model, with all scores reported as mean$\pm$SD over five runs (temperature$=$0). ASQA measures long-form ambiguous QA via QA-Hit. RewardBench evaluates preference alignment across four subsets: Chat, Chat Hard, Safety, and Reasoning. LongMemEval and LoCoMo assess long-term conversational memory, for which we report Contains Match and Accuracy, respectively, as primary semantic metrics.

On ASQA, \textit{\textbf{KG+HGT}} achieves the highest QA-Hit (83.2$\pm$1.7), outperforming both \textit{\textbf{Baseline}} (78.0$\pm$1.9) and \textit{\textbf{KG-only}} (81.8$\pm$1.8) on ambiguous multi-answer questions where focused context reduces hallucinated sub-answers. The gap widens as question ambiguity increases, with \textit{\textbf{KG+HGT}} maintaining consistency while unfiltered KG injection under \textit{\textbf{KG-only}} compounds errors (Table~\ref{tab:technical_eval}). RewardBench results show consistent gains across all four subsets (Table~\ref{tab:technical_eval}). \textit{\textbf{KG+HGT}} leads in every category: Chat (93.8$\pm$2.3), Chat Hard (91.6$\pm$2.2), Safety (96.3$\pm$2.1), and Reasoning (94.2$\pm$2.0). The largest improvements appear in Chat Hard (\textit{\textbf{KG+HGT}} 91.6$\pm$2.2 vs.\ \textit{\textbf{Baseline}} 68.4$\pm$2.4, $\Delta$23.2) and Reasoning (\textit{\textbf{KG+HGT}} 94.2$\pm$2.0 vs.\ \textit{\textbf{Baseline}} 71.8$\pm$2.2, $\Delta$22.4), where complex relational structure benefits most from selective grounding.

The ranking holds on the long-term memory benchmarks. On LongMemEval (Contains Match), \textit{\textbf{KG+HGT}} reaches 62.3$\pm$2.4 overall versus 57.0$\pm$0.8 (\textit{\textbf{KG-only}}) and 30.7$\pm$0.5 (\textit{\textbf{Baseline}}), and stays highest on every category (Temporal-reasoning 61.1$\pm$3.4, Multi-session 64.2$\pm$2.4). On LoCoMo (Accuracy), \textit{\textbf{KG+HGT}} leads across all subsets (Overall 88.7$\pm$1.2 vs.\ 86.1$\pm$1.5 and 84.2$\pm$0.8), with the biggest gains on Multi-hop (79.8$\pm$1.2 $\rightarrow$ 85.3$\pm$1.5) and Temporal (81.2$\pm$1.1 $\rightarrow$ 87.1$\pm$1.3), and the strongest adversarial robustness (92.8$\pm$1.0).

\begin{table}[!htbp]
\centering
\footnotesize
\setlength{\tabcolsep}{3pt}
\renewcommand{\arraystretch}{1.15}

\caption{Technical evaluation results. Values are mean$\pm$SD across 5 runs with fixed decoding (temperature$=$0). LongMemEval reports Contains Match (\%); LoCoMo reports Accuracy (\%).}
\label{tab:technical_eval}

\begin{tabular}{lp{1.8cm}ccc}
\toprule
& \textbf{Metric} &
\CB~\textit{\textbf{Baseline}} &
\CK~\textit{\textbf{KG-only}} &
\CH~\textit{\textbf{KG+HGT}}\\
\midrule

ASQA
& QA-Hit
& 78.0$\pm$1.9
& 81.8$\pm$1.8
& \textbf{83.2$\pm$1.7}\\

\cmidrule(lr){1-5}

\multirow{4}{*}{RewardBench}
& \cellcolor{gray!8}Chat
& \cellcolor{gray!8}72.3$\pm$2.1
& \cellcolor{gray!8}84.7$\pm$1.9
& \cellcolor{gray!8}\textbf{93.8$\pm$2.3}\\

& Chat Hard
& 68.4$\pm$2.4
& 71.2$\pm$2.0
& \textbf{91.6$\pm$2.2}\\

& \cellcolor{gray!8}Safety
& \cellcolor{gray!8}74.6$\pm$2.0
& \cellcolor{gray!8}88.1$\pm$1.8
& \cellcolor{gray!8}\textbf{96.3$\pm$2.1}\\

& Reasoning
& 71.8$\pm$2.2
& 85.4$\pm$2.1
& \textbf{94.2$\pm$2.0}\\

\cmidrule(lr){1-5}

\multirow{3}{*}{\shortstack{LongMem-\\Eval}}
& \cellcolor{gray!8}Overall
& \cellcolor{gray!8}30.7$\pm$0.5
& \cellcolor{gray!8}57.0$\pm$0.8
& \cellcolor{gray!8}\textbf{62.3$\pm$2.4}\\

& Temporal-reasoning
& 33.9$\pm$0.8
& 57.2$\pm$1.6
& \textbf{61.1$\pm$3.4}\\

& \cellcolor{gray!8}Multi-session
& \cellcolor{gray!8}25.8$\pm$1.2
& \cellcolor{gray!8}56.7$\pm$1.2
& \cellcolor{gray!8}\textbf{64.2$\pm$2.4}\\

\cmidrule(lr){1-5}

\multirow{6}{*}{LoCoMo}
& \cellcolor{gray!8}Overall
& \cellcolor{gray!8}84.2$\pm$0.8
& \cellcolor{gray!8}86.1$\pm$1.5
& \cellcolor{gray!8}\textbf{88.7$\pm$1.2}\\

& Single-hop
& 86.5$\pm$0.9
& 87.8$\pm$1.4
& \textbf{89.2$\pm$1.1}\\

& \cellcolor{gray!8}Multi-hop
& \cellcolor{gray!8}79.8$\pm$1.2
& \cellcolor{gray!8}82.5$\pm$1.8
& \cellcolor{gray!8}\textbf{85.3$\pm$1.5}\\

& Temporal
& 81.2$\pm$1.1
& 84.0$\pm$1.6
& \textbf{87.1$\pm$1.3}\\

& \cellcolor{gray!8}Open-domain
& \cellcolor{gray!8}85.0$\pm$1.0
& \cellcolor{gray!8}88.2$\pm$1.5
& \cellcolor{gray!8}\textbf{90.5$\pm$1.2}\\

& Adversarial
& 88.5$\pm$0.8
& 90.1$\pm$1.2
& \textbf{92.8$\pm$1.0}\\

\bottomrule
\end{tabular}
\end{table}

\paragraph{Takeaway.} These results confirm three points. First, selective entity grounding via HGT-based ranking outperforms exhaustive KG injection---\textit{\textbf{KG-only}} introduces noise that compounds errors on ambiguous queries, while \textit{\textbf{KG+HGT}} filters to structurally relevant context. Second, gains scale with task complexity: the more heterogeneous and ambiguous the query, the larger the improvement from selective grounding. Third, the ordering holds beyond the ambiguity and preference benchmarks: on two long-term memory benchmarks, selective grounding again outperforms both unfiltered graph memory and a strong LLM on the primary semantic metrics, indicating the advantage generalizes to sustained multi-session dialogue. This pattern motivates the framework's application to design conversation, where utterances are relationally dense and cognitively heterogeneous by nature.

\section{User Study}

Technical evaluation (Section~\ref{sec:4}) establishes that our pipeline produces measurably higher-quality responses than both ungrounded and na\"{i}vely grounded alternatives. To examine whether this technical advantage translates into designers' actual experience, we conducted a within-subjects study comparing three conditions---\CB~\textit{\textbf{Baseline}}, \CK~\textit{\textbf{KG-only}}, and \CH~\textit{\textbf{KG+HGT}}---across design tasks. In this study, we aimed to answer the following research questions:

\begin{itemize}
  \item[\textbf{RQ1:}] \textit{How does grounding LLM-based design conversation in a knowledge graph shape the system's ability to retain and leverage context across conversational turns?}
  \item[\textbf{RQ2:}] \textit{To what extent does a heterogeneous, personally constructed knowledge graph enable the system to interpret designer-specific language and intent beyond generic semantic processing?}
  \item[\textbf{RQ3:}] \textit{Does structurally grounded interaction reduce the cognitive effort required for designers to communicate their intent, and does it lead to deeper conversational engagement?}
\end{itemize}

\subsection{Study Design}

\paragraph{Participants.}
We recruited nine professional designers from diverse design domains---including graphic design, interior/furniture design, spatial design, store planning, exhibition/branding, product planning, and architecture (three males and six females; years of experience: M = 4.78; Appendix Table~\ref{tab:participants}). The study protocol was approved by our institutional ethics board. All participants reported active use of LLMs in their design workflow, rating their frequency of LLM use at 5.33 out of 7 ($SD = 1.66$) and their use of AI image generation tools at 5.00 out of 7 ($SD = 1.80$). This ensures that observed differences reflect knowledge-grounded interaction rather than novelty or unfamiliarity with conversational AI tools.

\paragraph{Conditions.}
The study employed a within-subjects design with three conditions, referred to throughout as \CB~\textit{\textbf{Baseline}}, \CK~\textit{\textbf{KG-only}}, and \textsc{\textbf{\textcolor{cCog}{CogChat}}}~(\CH, \textit{\textbf{KG+HGT}}). In \textit{\textbf{Baseline}}, participants interacted with the LLM without any external knowledge grounding. In \textit{\textbf{KG-only}}, the LLM was augmented with a personal knowledge graph constructed in real time but without embedding-based selection. In \textsc{\textbf{\textcolor{cCog}{CogChat}}}, the system ran the full pipeline---graph construction, HGT embedding, selective entity injection, and distribution-based probing. The order of conditions was counterbalanced across participants.

\paragraph{Tasks.}
Each participant completed three design tasks, one per condition:\textbf{ (T1)} Design a modular urban micro-mobility rental station that prioritizes space efficiency; \textbf{(T2)} Design a modern and minimal lamp with personal preferences; and \textbf{(T3)} Design a visual identity for an eco-friendly camping gear brand. Task--condition assignment was counterbalanced alongside condition order to prevent confounding between specific tasks and specific systems.

\paragraph{Procedure.}
Each session lasted approximately 120 minutes per participant. After a 5-minute briefing explaining the study purpose and interface, participants completed three consecutive system sessions, each lasting up to 30 minutes followed by a 5-minute post-condition survey. Participants were free to end a session early if they felt they had reached a satisfactory design outcome. After all three conditions, a 15-minute semi-structured interview was conducted in which participants compared their experiences across conditions, described their perception of conversational depth and personalization, and reflected on the role of probing questions in their design process.

\paragraph{Measures.}
We collected both quantitative and qualitative data. Self-report measures administered after each condition included the Chatbot Usability Questionnaire~\cite{holmes2019usability} (CUQ; 16 items, 5-point scale), NASA-TLX~\cite{hart1988development}, Intent Expression~\cite{venkatesh2008technology}, Satisfaction \& Output Quality~\cite{kang2021metamap}, and Agency/Ownership~\cite{chan2022investigating}. Automated conversational metrics were computed from interaction logs: Semantic Similarity (BERTScore~\cite{zhang2019bertscore}, per-turn) and Token-level IoU for lexical overlap. Behavioral measures included Task Completion Time (TCT), total turns, number of conversation sessions, and Probing Success Rate (the proportion of system-generated questions that participants chose to answer). Finally, semi-structured interviews were transcribed and analyzed through thematic analysis~\cite{braun2006using} to identify experiential patterns across conditions. 
\paragraph{Analysis.}
Given the within-subjects design with three conditions and a small sample ($n = 9$), we used Friedman tests for omnibus comparisons. Where the omnibus test was significant, pairwise Wilcoxon signed-rank tests were conducted with Benjamini--Hochberg correction for multiple comparisons~\cite{benjamini1995controlling}. Kendall's $W$ is reported as effect size, with $W > .50$ considered large~\cite{tomczak2014need}.

\section{Results \& Findings}
\label{sec:6}
\begin{figure}[!htbp]
  \centering
  \includegraphics[width=\linewidth]{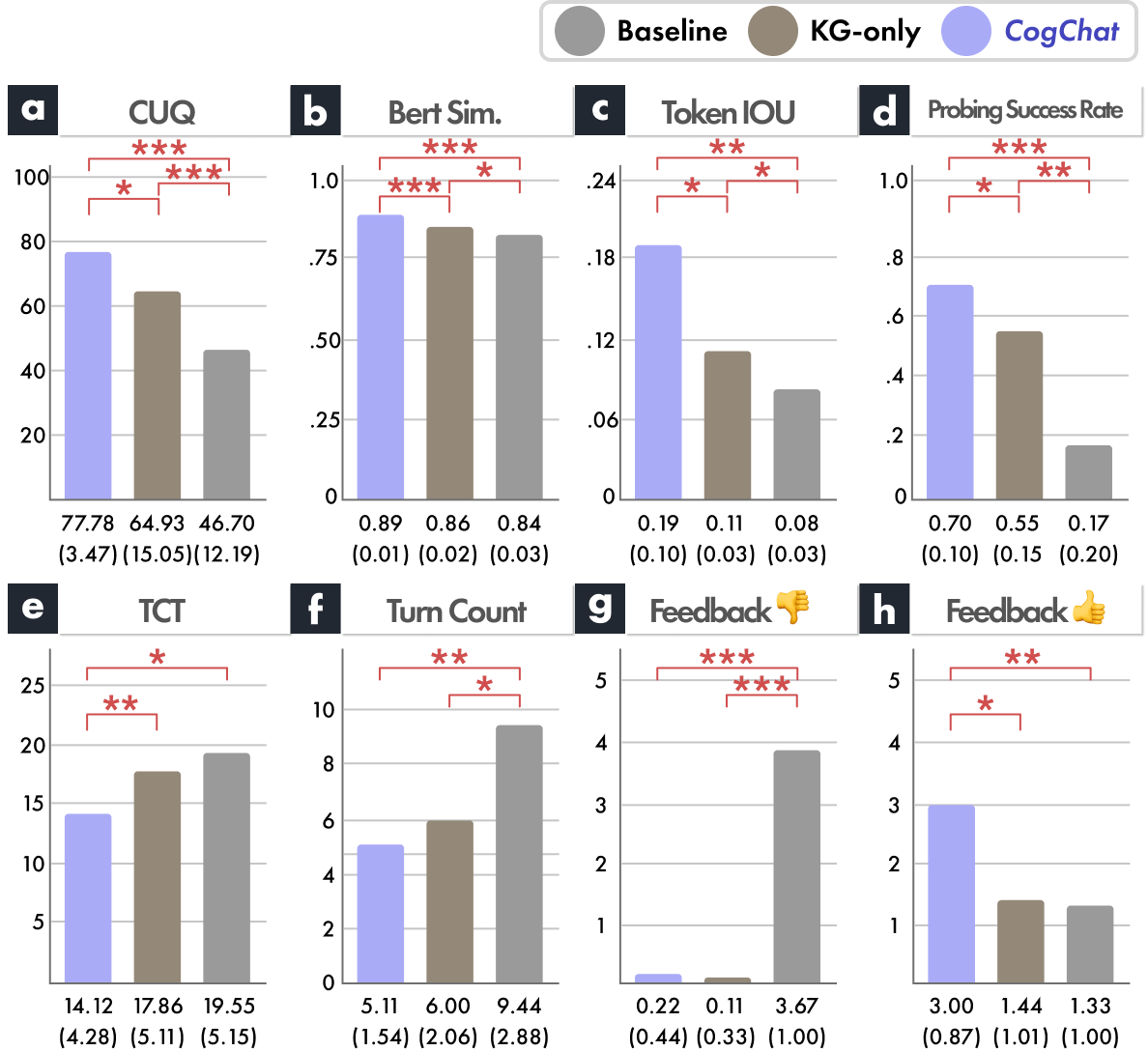}
  \caption{Quantitative results across three conditions. Top row: (a) CUQ, (b) semantic similarity (BERTScore, per-turn), (c) token-level IoU, (d) probing success rate. Bottom row: (e) task completion time, (f) turn count, (g) negative feedback count, (h) positive feedback count. Significance brackets show pairwise Wilcoxon signed-rank tests with Benjamini--Hochberg correction (* $p<.05$, ** $p<.01$, *** $p<.001$).}
  \Description{Eight-panel bar chart. Top row: (a) CUQ scores with CogChat highest at 77.8, (b) BERTScore semantic similarity with CogChat at .887, (c) Token-level IoU with CogChat at .185, (d) Probing success rate with CogChat at .70. Bottom row: (e) Task completion time with CogChat lowest at 14.1 min, (f) Turn count with CogChat lowest at 5.1, (g) Negative feedback with Baseline highest at 3.67, (h) Positive feedback with CogChat highest at 3.00. Significance brackets show pairwise Wilcoxon signed-rank tests.}
  \label{fig:quant}
\end{figure}

\begin{figure*}[!htbp]
  \centering
  \includegraphics[width=\textwidth]{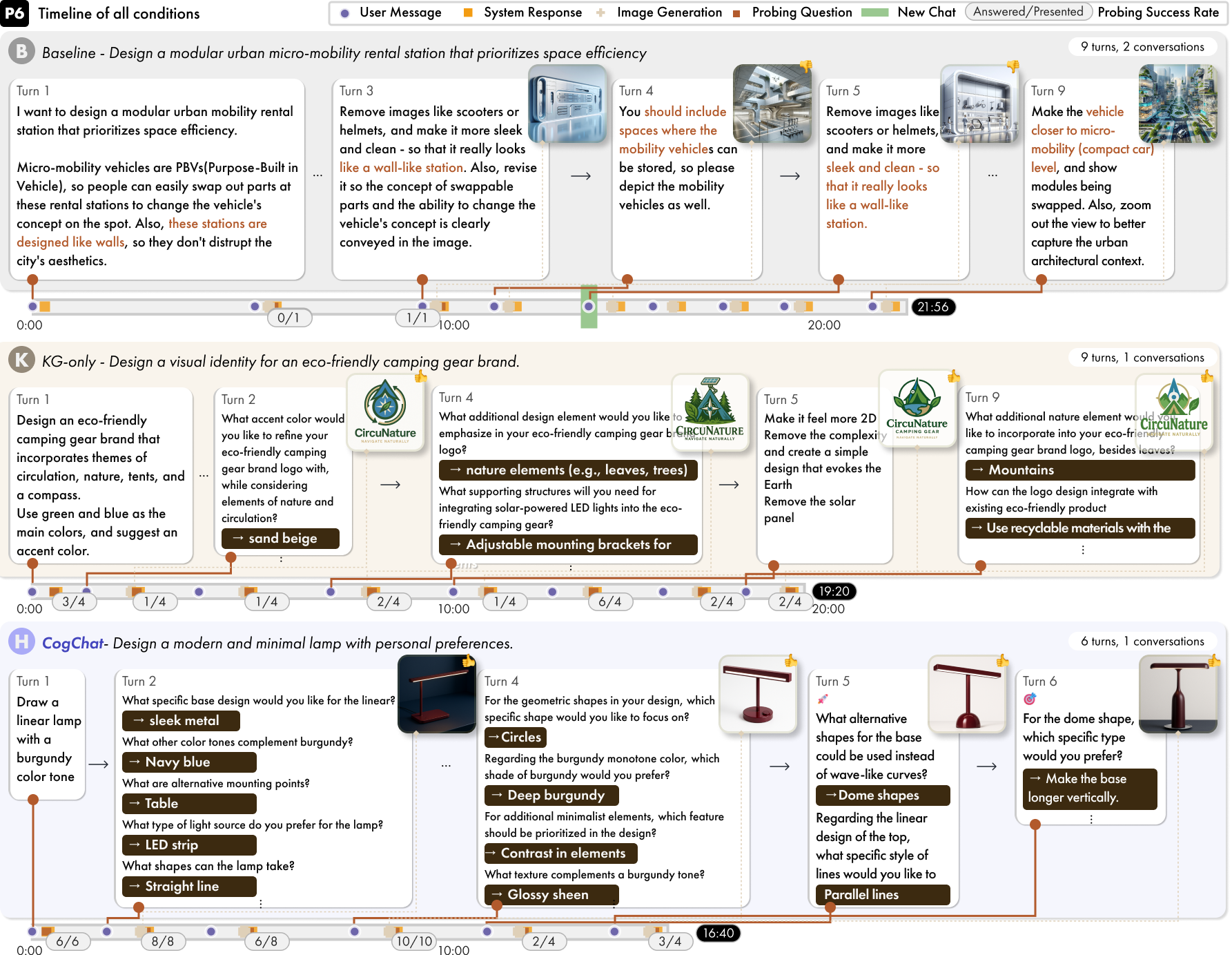}
  \caption{P6's interaction timeline across three conditions.  \textit{\textbf{Baseline}} (top) required 2 conversations with repeated re-explanations across 9 turns. \textit{\textbf{KG-only}}~(middle) maintained context in a single session but produced redundant questions. \textsc{\textbf{\textcolor{cCog}{CogChat}}} (bottom) completed the task in 6 turns with progressively specific probing questions and the highest probing success rate.}
  \Description{Three-panel interaction timeline for participant P6. The top panel shows the Baseline condition with two separate conversation sessions and 9 turns involving repeated re-explanations. The middle panel shows the KG-only condition maintaining a single session but with redundant probing questions. The bottom panel shows the CogChat condition completing the task in 6 turns with progressively specific questions and the highest probing success rate.}
  \label{fig:session}
\end{figure*}

\subsection{RQ1: Context Retention Across Conversational Turns}

Turn count decreased significantly across conditions, with KG-grounded conditions requiring fewer turns to reach a satisfactory outcome. \textsc{\textbf{\textcolor{cCog}{CogChat}}} showed an average of 5.1 turns, \textit{\textbf{KG-only}} 6.0, and \textit{\textbf{Baseline}} 9.4 ($\chi^2(2) = 14.35$, $W = .80$, $p < .001$; Figure~\ref{fig:quant}f). \textsc{\textbf{\textcolor{cCog}{CogChat}}} showed significantly fewer turns than \textit{\textbf{Baseline}} ($p < .001$) with 45.9\% reduction, and \textit{\textbf{KG-only}} showed significantly fewer than \textit{\textbf{Baseline}} ($p < .05$) with 36.4\% reduction. Conversation count reinforced this: \textit{\textbf{Baseline}} required 2.1 sessions on average, while both \textit{\textbf{KG-only}} (1.1) and \textsc{\textbf{\textcolor{cCog}{CogChat}}} (1.0) completed tasks in a single session ($\chi^2(2) = 12.29$, $W = .68$, $p = .002$)---designers stopped restarting once the system retained their prior input. Negative feedback further isolated the source of \textit{\textbf{Baseline}}'s failure: 3.67 dislikes per session versus 0.11 for \textit{\textbf{KG-only}} and 0.22 for \textsc{\textbf{\textcolor{cCog}{CogChat}}} ($\chi^2(2) = 16.27$, $W = .90$, $p < .001$; Figure~\ref{fig:quant}g), with no difference between KG-grounded conditions---indicating that context loss, not response quality per se, drove designer dissatisfaction.

Within the KG-grounded conditions, HGT contributed finer-grained context retention. Semantic similarity (BERTScore, per-turn) differed significantly across all three pairs ($\chi^2(2) = 14.89$, $W = .83$, $p < .001$), with \textsc{\textbf{\textcolor{cCog}{CogChat}}} achieving .887, \textit{\textbf{KG-only}} .862, and \textit{\textbf{Baseline}} .844 (Figure~\ref{fig:quant}b). Crucially, the \textit{\textbf{KG-only}} vs.\ \textsc{\textbf{\textcolor{cCog}{CogChat}}} gap was itself significant, indicating that retaining entities alone is insufficient. Token-level IoU confirmed this at the lexical level ($\chi^2(2) = 16.22$, $W = .90$, $p < .001$): \textsc{\textbf{\textcolor{cCog}{CogChat}}} (.185) nearly doubled \textit{\textbf{KG-only}} (.110) and more than doubled \textit{\textbf{Baseline}} (.077) (Figure~\ref{fig:quant}c), suggesting that HGT embedding enables the system to reuse the designer's own vocabulary rather than paraphrasing generically. Notably, \textit{\textbf{KG-only}} and \textsc{\textbf{\textcolor{cCog}{CogChat}}} accumulated comparable graphs by session end---80.12 ($SD = 36.48$) and 88.33 ($SD = 28.85$) nodes; 129.75 ($SD = 69.69$) and 130.00 ($SD = 48.26$) edges---yet both selected subgraphs of similar size per turn: 22.96 ($SD = 2.84$) vs.\ 21.52 ($SD = 2.70$) nodes, 27.29 ($SD = 2.86$) vs.\ 26.24 ($SD = 2.88$) edges. The difference lies in selectivity: per-participant node utilization averaged 43.9\% for \textit{\textbf{KG-only}} but only 26.8\% for \textsc{\textbf{\textcolor{cCog}{CogChat}}}, indicating that selective grounding draws more precisely from a richer pool rather than consuming more context.

In interviews, the contrast was consistent. In \textit{\textbf{Baseline}}, P2 observed that ``only the most recent request was reflected, or parts dropped entirely,'' P5 described outputs ``getting worse to the point where the first image was better,'' and P1 reported opening new sessions out of frustration. In \textit{\textbf{KG-only}}, P6 noted it ``grasped intent well based on prior conversation,'' but P3 and P5 reported repeated questions about already-established values (e.g., re-asking about 4000K LED). In \textsc{\textbf{\textcolor{cCog}{CogChat}}}, P2 reported that ``the intent was preserved throughout and subsequent requests reflected accurately,'' and P5 noted ``responses became more precise as the conversation went on.'' Figure~\ref{fig:session} illustrates this pattern for P6, whose \textit{\textbf{Baseline}}~session required two conversations with repeated re-explanations, while \textsc{\textbf{\textcolor{cCog}{CogChat}}}~completed the task in a single session with 6 turns.

These findings suggest that KG structure provides the foundation for context retention, while HGT embedding refines it---preserving not just which concepts were mentioned but how they relate to each other across turns.

\begin{figure*}[!htbp]
  \centering
  \includegraphics[width=\textwidth]{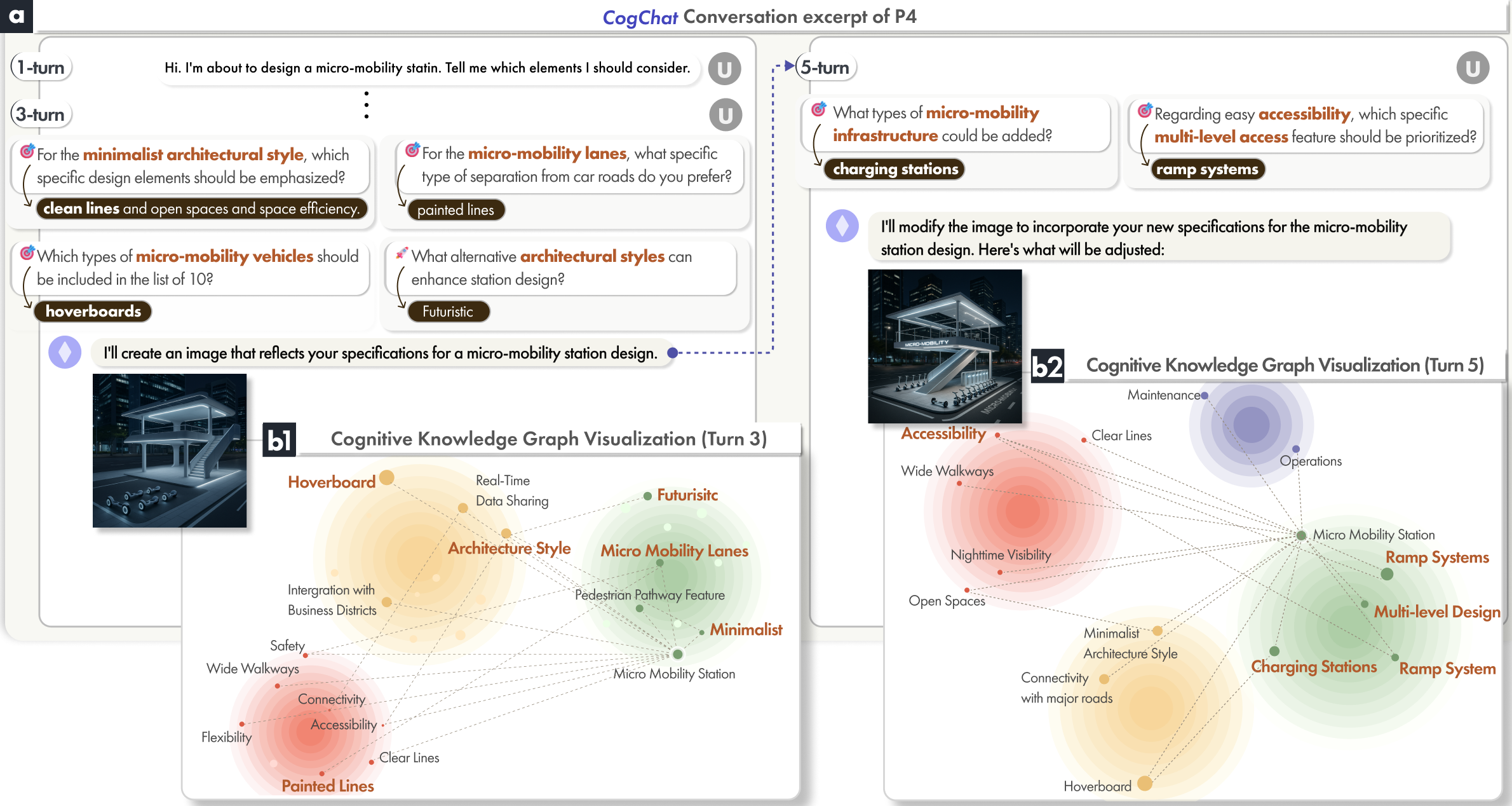}
  \caption{(a) P4's conversation excerpt with \textsc{\textbf{\textcolor{cCog}{CogChat}}} and cognitive knowledge graph snapshots at turn 3 (b1) and turn 5 (b2).\textit{ Intentional questions} narrowed design specifications; an \textit{exploratory question} introduced ``futuristic'' as an alternative direction. The graph densifies as probing responses accumulate across turns.}
  \Description{Two-panel figure for participant P4. Panel (a) shows a conversation excerpt with CogChat including intentional and exploratory probing questions. Panel (b) shows cognitive knowledge graph snapshots at turn 3 and turn 5, illustrating graph densification as probing responses accumulate, with an exploratory node labeled futuristic highlighted in purple.}
  \label{fig:probing}
\end{figure*}

\subsection{RQ2: Personalized Interpretation of Designer-Specific Language}
CUQ scores showed significant differences across all three pairs, with \textsc{\textbf{\textcolor{cCog}{CogChat}}} scoring 77.8, \textit{\textbf{KG-only}} 64.9, and \textit{\textbf{Baseline}} 46.7 ($\chi^2(2) = 16.22$, $W = .90$, $p < .001$; Figure~\ref{fig:quant}a). Notably, the \textit{\textbf{KG-only}} vs.\ \textsc{\textbf{\textcolor{cCog}{CogChat}}} gap was itself significant---retaining a knowledge graph improved usability, but embedding-based selection added a further perceptible layer of understanding. Intent Expression items I2 (``I was able to accurately express my intent'') and I3 (``the system understood my intent'') were significant across all three pairwise comparisons ($\chi^2(2) = 13.24$, $W = .74$ and $13.61$, $W = .76$, both $p < .002$; Appendix Figure~\ref{fig:survey}, I2--I3), suggesting that \textsc{\textbf{\textcolor{cCog}{CogChat}}} did not merely remember what designers said but resolved what they meant.

Probing Success Rate---the proportion of system-generated questions that participants chose to answer---differed significantly across all three pairs ($\chi^2(2) = 12.67$, $W = .70$, $p = .002$; Figure~\ref{fig:quant}d). \textsc{\textbf{\textcolor{cCog}{CogChat}}} achieved a rate of .70, \textit{\textbf{KG-only}} .55, and \textit{\textbf{Baseline}} .17. Since designers could freely skip questions, this behavioral measure indicates that \textsc{\textbf{\textcolor{cCog}{CogChat}}}'s graph-grounded questions were perceived as nearly four times more relevant than \textit{\textbf{Baseline}}'s generic follow-ups. Output Quality items O1, O3, O4, and O5 showed three-way significance, with \textsc{\textbf{\textcolor{cCog}{CogChat}}} consistently highest (Appendix Figure~\ref{fig:survey}, O1--O5). Positive feedback reinforced this asymmetry: \textsc{\textbf{\textcolor{cCog}{CogChat}}} averaged 3.00 likes compared to \textit{\textbf{KG-only}} 1.44 and \textit{\textbf{Baseline}} 1.33 ($\chi^2(2) = 10.90$, $W = .61$, $p = .004$; Figure~\ref{fig:quant}h), with \textit{\textbf{Baseline}} vs.\ \textit{\textbf{KG-only}} not significant---confirming that positive engagement tracks HGT-based grounding, not KG presence alone.

In interviews, the quality gap between \textit{\textbf{KG-only}} and \textsc{\textbf{\textcolor{cCog}{CogChat}}} questions was a recurring theme. P3 reported that \textit{\textbf{KG-only}} ``asked about values I had already set and gave answers that just flipped the same shape left and right.'' In contrast, P5 described \textsc{\textbf{\textcolor{cCog}{CogChat}}} as presenting specific design elements---font choices, color textures, compositional details---as concrete options rather than generic suggestions, calling the experience ``the most joyful.'' P8 found that the system interpreted ``water properties'' not as transparency but as gentle wave texture---``actually better---I received design advice in reverse with \includegraphics[height=1em]{Figure/exp.png} \textit{exploratory questions}.'' P6 stated that \textsc{\textbf{\textcolor{cCog}{CogChat}}} questions helped ``narrow down ideas much more smoothly.'' In \textit{\textbf{Baseline}}, P5 summarized the difficulty: ``I had to communicate everything through my own language alone.'' Figure~\ref{fig:probing} shows how P4's \includegraphics[height=1em]{Figure/intent.png} \textit{intentional questions} progressively narrowed design specifications across turns, while an \includegraphics[height=1em]{Figure/exp.png} \textit{exploratory question} introduced ``futuristic'' as an unexpected direction. The KG visualization at turn 3 (Figure~\ref{fig:probing}$b_1$) captures initial concepts such as Minimalist and Micro Mobility Lanes; by turn 5 (Figure~\ref{fig:probing}$b_2$), probing responses have added Charging Stations, Ramp Systems, and Multi-level Design, illustrating how the graph densifies through the feedback loop.

These patterns indicate that personalization requires not just retaining a designer's vocabulary (RQ1) but structuring the relations among their concepts.

\subsection{RQ3: Cognitive Effort and Conversational Depth}
NASA-TLX scores showed significantly lower cognitive load in KG-grounded conditions (Appendix Figure~\ref{fig:survey}, N1--N6). Mental Demand ($\chi^2(2) = 8.36$, $W = .46$, $p = .015$), Effort ($\chi^2(2) = 8.42$, $W = .47$, $p = .015$), and Frustration ($\chi^2(2) = 13.00$, $W = .72$, $p = .002$) all followed the pattern \textit{\textbf{Baseline}} $>$ \textit{\textbf{KG-only}} $>$ \textsc{\textbf{\textcolor{cCog}{CogChat}}}, with \textit{\textbf{Baseline}} vs.\ \textit{\textbf{KG-only}} and \textit{\textbf{Baseline}} vs.\ \textsc{\textbf{\textcolor{cCog}{CogChat}}} significant but \textit{\textbf{KG-only}} vs.\ \textsc{\textbf{\textcolor{cCog}{CogChat}}} not. This suggests that the context retention established in RQ1 is the primary driver of cognitive load reduction---once the system stops losing context, designers stop spending effort re-explaining.

Task completion time showed a different pattern. \textsc{\textbf{\textcolor{cCog}{CogChat}}} averaged 14.1 min, \textit{\textbf{KG-only}} 17.9 min, and \textit{\textbf{Baseline}} 19.5 min ($\chi^2(2) = 8.22$, $W = .46$, $p = .016$; Figure~\ref{fig:quant}e). \textsc{\textbf{\textcolor{cCog}{CogChat}}} was significantly faster than both \textit{\textbf{Baseline}} ($p < .05$) with 27.7\% reduction and \textit{\textbf{KG-only}} ($p < .01$) with 21.0\% reduction, while \textit{\textbf{Baseline}} vs.\ \textit{\textbf{KG-only}} was not significant. Context retention alone (\textit{\textbf{KG-only}}) does not reduce task time; the efficiency gain requires the personalized grounding of \textsc{\textbf{\textcolor{cCog}{CogChat}}}. Satisfaction items S2 ($\chi^2(2) = 13.61$, $W = .76$, $p = .001$) and S3 ($\chi^2(2) = 12.07$, $W = .67$, $p = .002$) showed the same \textit{\textbf{KG-only}} vs.\ \textsc{\textbf{\textcolor{cCog}{CogChat}}} pattern (Appendix Figure~\ref{fig:survey}, S2--S3), reinforcing that personalized interpretation converts retained context into faster, more satisfying outcomes.

Agency and Ownership showed three-way significance ($\chi^2(2) = 12.25$, $W = .68$, $p = .002$), with \textsc{\textbf{\textcolor{cCog}{CogChat}}} rated highest (Appendix Figure~\ref{fig:survey}, A2). P8 distinguished ``the AI helped me'' from ``this feels like my design,'' noting that \textsc{\textbf{\textcolor{cCog}{CogChat}}} achieved the latter when the designer maintained directional initiative. Several participants also remarked that double-clicking a word in chat to inspect its connections made the system's understanding tangible; P5 noted ``I could see my own logic laid out, so I trusted the responses more,'' and P3 added that the structured view confirmed her intent had been preserved rather than diluted across turns. Regarding depth, 8 of 9 participants identified \textsc{\textbf{\textcolor{cCog}{CogChat}}} as the deepest conversation. Definitions clustered around three themes: specificity of response (P3, P5-8), multi-perspectival engagement (P2, P4), and progressive refinement (P1). P9 uniquely identified \textit{\textbf{KG-only}} as deepest, defining depth as creative divergence, and proposed using \textsc{\textbf{\textcolor{cCog}{CogChat}}} to concretize intent first, then \textit{\textbf{KG-only}} to diversify. 

Together, context retention (RQ1) reduces the cognitive burden, and personalized grounding (RQ2) converts that relief into efficiency and depth---\textsc{\textbf{\textcolor{cCog}{CogChat}}} amplifies the designer's reasoning rather than replacing it.

\section{Discussion \& Limitation}
Our findings show that preserving relational context requires structure, not just capacity. Recency-based systems lose cross-turn relations regardless of window size; our \textit{\textbf{Baseline}} confirms this, as it received the full conversation history in the context window. The performance gap thus reflects not information absence but the difference between having context and structuring it. A knowledge graph retains cross-turn relations, but only HGT-based selection makes that retention useful. The gap between \textit{\textbf{KG-only}} and \textsc{\textbf{\textcolor{cCog}{CogChat}}} reveals that unfiltered graph injection introduces noise that competes with salient nodes for LLM attention---more context degrades rather than improves output when relevance is not enforced. This suggests that the bottleneck in LLM-based design conversation is not memory capacity but context curation: deciding which relations to foreground at each turn. Our three-condition ablation separates general graph memory (\textit{\textbf{KG-only}}) from selective heterogeneous grounding (\textsc{\textbf{\textcolor{cCog}{CogChat}}}). Graph memory alone handles simple retention: \textit{\textbf{KG-only}} and \textsc{\textbf{\textcolor{cCog}{CogChat}}} match on turn count and frustration. HGT drives the gains on ambiguity resolution, probing success, semantic similarity, intent interpretation, and efficiency. We therefore claim no universal superiority---the benefits concentrate on relationally dense, ambiguity-heavy conversation, while \textit{\textbf{KG-only}} often suffices for simple retention.

We frame \textsc{\textbf{\textcolor{cCog}{CogChat}}}'s contribution as graph-grounded conversational personalization through structured relational memory---reusing the relations a designer expresses rather than modeling their internal cognition. Grounding conversational AI in a personal heterogeneous knowledge graph enhances the design process. For design specifically, where ``warm'' can index tactile, chromatic, and emotional dimensions at once, context selection must be type-aware---precisely what HGT's heterogeneous attention provides. Beyond response generation, probing questions functioned as a \textbf{cognitive scaffold}---clarifying ambiguities while prompting designers to consider adjacent conceptual territories, reducing cognitive load and deepening engagement. Our results highlight that what matters is not what the system remembers but how it organizes what it remembers---relational structure, not vocabulary, drives the observed gains.

Despite these promising results, our approach has several limitations. First, on the graph side, knowledge graph extraction relies on the capabilities of the underlying LLM (GraphRAG via GPT-4o). In highly abstract or purely visual design discussions, text-based entity extraction may fail to capture nuanced visual intents, and future work should explore multimodal knowledge graphs that directly embed visual elements alongside textual nodes. Extraction errors, however, are unlikely to propagate. Extraction itself is largely accurate---88.6\% precision across 1,226 entities and 1,710 relations under strict LLM-based judgment---and the graph is re-embedded each turn, so weakly supported nodes become progressively less likely to be selected. The consistent \textsc{\textbf{\textcolor{cCog}{CogChat}}} $>$ \textit{\textbf{KG-only}} gap indicates HGT suppresses graph noise rather than amplifying it, and in-context inspection lets designers verify that the graph matches their intent. The graph also has a cold-start limitation: when only a few concepts have been expressed, \textsc{\textbf{\textcolor{cCog}{CogChat}}}'s advantage over a strong LLM---one that asks clarifying questions and maintains a preference summary---is small, and grows only as relational structure accumulates.

Second, on the evaluation side, our study---nine expert designers on short, single-session tasks---should not be read as evidence of long-term personalization; although effect sizes were large ($W > 0.7$), future longitudinal studies are needed to track how personal knowledge graphs evolve over extended, multi-session projects. Additionally, managing graph decay---pruning outdated nodes, resolving contradictory constraints, and handling intent pivots---remains an open challenge for long-term deployment. Our evaluation also centers on process, perception, and self-reported output quality rather than a blind assessment of the generated artifacts; blind expert ratings of final designs, or a qualitative cross-condition comparison of outcomes, are an important next step for claims about design generation. Finally, because probing and preference organization are partly shared across conditions, fully disentangling their contributions from heterogeneous graph grounding requires targeted ablations that we leave to future work.

\section{Conclusion}

We introduced \textsc{\textbf{\textcolor{cCog}{CogChat}}}, a novel real-time conversational framework that grounds LLM interaction in a personal heterogeneous knowledge graph constructed dynamically from a designer's input. By addressing the structural limitations of generic recency-based context, \textsc{\textbf{\textcolor{cCog}{CogChat}}} prevents relational context decay, resolves personalized semantics, and drives deeper conversational engagement. Technical evaluations confirmed that HGT-based entity selection outperforms ungrounded and na\"{i}vely grounded models by filtering noise and injecting highly relevant structural context. A within-subjects study with professional designers further validated that our system significantly improves context retention, enhances personalized interpretation, and reduces the cognitive effort required to communicate design intent. These findings suggest that structuring a designer's expressed concepts and relations as a dynamic knowledge graph can preserve relational context that fades across turns, enabling conversational AI to ground interaction in how designers \textit{think} rather than what they said most recently.

\begin{acks}
This work was supported by the Industrial Technology Innovation Program(RS-2025-02317326, Development of AI-Driven Design Generation Technology Based on Designer Intent) funded by the Ministry of Trade, Industry \& Energy(MOTIE, Korea)
\end{acks}

\bibliographystyle{ACM-Reference-Format}
\bibliography{reference}

\appendix
\section*{Appendices}
\section{System Implementation}
\subsection{Probing Question Generation Prompts}
\label{app:1}
Each condition generates intentional and exploratory probing questions through distinct prompting strategies. \textit{\textbf{Baseline}} uses a single unified prompt with no external context. \textit{\textbf{KG-only}} leverages graph connectivity (1--3 hop neighbors) and prior answers. \textsc{\textbf{\textcolor{cCog}{CogChat}}} uses HGT embedding similarity to resolve ambiguity and expand into semantically adjacent concepts.
\subsubsection{\textbf{KG-only}}
The KG-only condition generates probing questions by traversing graph structure. Intentional questions draw from 1--2 hop connections with reference to previously answered questions; exploratory questions reach 2--3 hops to surface creative alternatives.

\noindent{\sffamily\small\bfseries Intentional probing prompt (KG-only)}
\begin{lstlisting}[style=promptK]
User said: "{user_message}"
Related concepts from knowledge graph (1-2 hop connections):
{intentional_context}
PREVIOUSLY ANSWERED QUESTIONS:
{previously_answered}
Generate {num_intentional} questions that:
1. Use 1-hop and 2-hop graph connections to suggest related options
2. Build on previous choices to explore variations
Return JSON format:
{
  "questions": [
    {
      "question": "specific clarifying question to resolve ambiguity",
      "suggested_answers": ["option 1", "option 2", "option 3"]
    }
  ]
}
\end{lstlisting}

\noindent{\sffamily\small\bfseries Exploratory probing prompt (KG-only)}
\begin{lstlisting}[style=promptK]
User said: "{user_message}"
Related concepts from knowledge graph (2-3 hop connections):
{exploratory_context}
Generate {num_exploratory} questions that:
1. Use 2-3 hop graph connections to find creative alternatives
2. Suggest complementary or contrasting options
3. Expand into related but different directions
CREATIVE EXPLORATION RULES:
- Look beyond direct connections to find interesting combinations
- Introduce related concepts from the broader graph
Return JSON format:
{
  "questions": [
    {
      "question": "creative exploration question using graph",
      "suggested_answers": ["option 1", "option 2", "option 3"]
    }
  ]
}
\end{lstlisting}

\subsubsection{ \textsc{\textbf{\textcolor{cCog}{CogChat}}}}
The \textbf{\textit{KG+HGT}} condition replaces structural traversal with embedding-based selection. \includegraphics[height=1em]{Figure/intent.png}~Intentional questions target high-similarity but underspecified nodes to resolve ambiguity; \includegraphics[height=1em]{Figure/exp.png}~exploratory questions sample from the broader embedding space to surface unexpected connections.

\noindent{\sffamily\small\bfseries Intentional probing prompt (KG+HGT)}
\begin{lstlisting}[style=promptH]
User said: "{user_message}"
High-importance concepts from HGT embeddings (semantic similarity-weighted):
{intentional_context}
PREVIOUSLY ANSWERED:
{previously_answered}
Generate {num_intentional} CLARIFYING questions that:
1. Transform abstract concepts into concrete options
2. Use HGT embeddings to find semantically related but more specific options
Return JSON format:
{
  "questions": [
    {
      "question": "specific clarifying question to resolve ambiguity",
      "suggested_answers": ["option 1", "option 2", "option 3"]
    }
  ]
}
\end{lstlisting}

\noindent{\sffamily\small\bfseries Exploratory probing prompt (KG+HGT)}
\begin{lstlisting}[style=promptH]
User said: "{user_message}"
Semantically-related concepts from HGT embeddings (broader exploration space):
{exploratory_context}
Generate {num_exploratory} CREATIVE questions that:
1. Use semantic embeddings to discover unexpected but relevant connections
2. Go beyond direct graph connections to find creative alternatives
CREATIVE EXPANSION RULES:
- Use semantic similarity to find concepts that "feel related"
- Balance novelty with relevance to the conversation
Return JSON format:
{
  "questions": [
    {
      "question": "creative exploratory question using semantic similarity",
      "suggested_answers": ["option 1", "alternative 2", "option 3"]
    }
  ]
}
\end{lstlisting}

\subsection{Response Generation Prompts}
\label{app:2}
Response generation prompts differ in the context injected into the LLM. \textit{\textbf{Baseline}} receives the full conversation history. \textit{\textbf{KG-only}} receives the most recent graph nodes. \textsc{\textbf{\textcolor{cCog}{CogChat}}} receives the top-$k$ nodes ranked by HGT embedding similarity.
\subsubsection{\textbf{Baseline}}\leavevmode\par\vspace{1pt}
\noindent{\sffamily\small\bfseries Response generation prompt (Baseline, LLM-only)}
\begin{lstlisting}[style=promptB]
You are a helpful AI assistant.
Maintain continuity with the ongoing conversation.
Conversation History:
{context_history}
Instructions:
- Preserve prior user decisions, preferences, and constraints.
- If the current message refers implicitly to earlier content,
  resolve it from the history.
- Do not restart the topic unless the user clearly changes it.
- Always respond in the same language as the user's message.
\end{lstlisting}

\subsubsection{\textbf{KG-only}}
\noindent{\sffamily\small\bfseries Response generation prompt (KG-only)}
\begin{lstlisting}[style=promptK]
You are a helpful design assistant with access to a knowledge graph.
Current knowledge graph contains:
{context_str}
Note: {context_str} contains recent nodes from the graph:
- node_1 (type): description
- node_2 (type): description
...
Use this context to provide informed responses about design concepts,
properties, and relationships. Maintain conversation continuity by
referencing recent topics when appropriate.
\end{lstlisting}

\subsubsection{ \textsc{\textbf{\textcolor{cCog}{CogChat}}}} \leavevmode\par\vspace{1pt}
\noindent{\sffamily\small\bfseries Response generation prompt (KG+HGT)}
\begin{lstlisting}[style=promptH]
You are a helpful design assistant with access to a knowledge graph.
Current knowledge graph contains (HGT-selected by semantic similarity):
{context_str}
Note: {context_str} contains Top-k nodes selected by HGT embedding similarity:
- node_1 (type): description  [similarity: 0.87]
- node_2 (type): description  [similarity: 0.82]
...
\end{lstlisting}

\subsection{Image Generation Prompts}
\label{app:3}
Image generation prompts share the same three-tier context structure as response generation. \textit{\textbf{Baseline}} relies on conversation history. \textit{\textbf{KG-only}} appends recent graph nodes as design context. \textsc{\textbf{\textcolor{cCog}{CogChat}}} injects HGT-ranked nodes with similarity scores, grounding the visual output in the designer's accumulated conceptual structure.
\subsubsection{\textbf{Baseline}}\leavevmode\par\vspace{1pt}
\noindent{\sffamily\small\bfseries Image generation prompt (Baseline, LLM-only)}
\begin{lstlisting}[style=promptB]
You are helping with an ongoing conversation. The user has requested a new image.
IMPORTANT CONTEXT:
[History context with user Q&A]
Previously generated image:
{last_image_prompt}
USER REQUEST: "{message}"
Create a detailed image generation prompt that:
1. MUST include ALL user choices from Q&A (colors, styles, materials, etc.)
2. Uses the conversation context to understand what the user wants
3. Creates a cohesive prompt that combines all their stated preferences
Return ONLY the image generation prompt, nothing else.
\end{lstlisting}

\subsubsection{\textbf{KG-only}}
\noindent{\sffamily\small\bfseries Image generation prompt (KG-only)}
\begin{lstlisting}[style=promptK]
You are helping with an ongoing conversation. The user has requested a new image.
IMPORTANT CONTEXT:
[History context with user Q&A]
Knowledge graph context (related design concepts):
- modern_chair (furniture): Sleek minimalist design with chrome legs
- steel_finish (material): Brushed steel with matte coating
- red_accent (color): Vibrant red upholstery fabric
- compact_size (property): Space-efficient 80cm width
[... more graph nodes ...]
Previously generated image:
{last_image_prompt}
USER REQUEST: "{message}"
TASK:
Create a detailed image generation prompt that:
1. MUST include ALL user choices from Q&A (colors, styles, materials, etc.)
2. Incorporates knowledge graph concepts that are relevant
3. Combines user preferences with graph-based design knowledge
Return ONLY the image generation prompt, nothing else.
\end{lstlisting}

\subsubsection{ \textsc{\textbf{\textcolor{cCog}{CogChat}}}}
\leavevmode\par\vspace{1pt}
\noindent{\sffamily\small\bfseries Image generation prompt (KG+HGT)}
\begin{lstlisting}[style=promptH]
You are helping with an ongoing conversation. The user has requested a new image.
IMPORTANT CONTEXT:
[History context with user Q&A]
Knowledge graph context (HGT-selected by semantic similarity):
- modern_chair (furniture): Sleek minimalist design [similarity: 0.89]
- steel_finish (material): Brushed steel coating [similarity: 0.84]
- ergonomic_design (property): Lumbar support feature [similarity: 0.81]
[... more semantically related nodes ...]
Previously generated image:
{last_image_prompt}
USER REQUEST: "{message}"
TASK:
Create a detailed image generation prompt that:
1. MUST include ALL user choices from Q&A (colors, styles, materials, etc.)
2. Incorporates knowledge graph concepts that are relevant
3. Combines user preferences with graph-based design knowledge
Return ONLY the image generation prompt, nothing else.
\end{lstlisting}

\section{Participant Demographics and Survey Instruments}
\renewcommand{\thetable}{B\arabic{table}}
\setcounter{table}{0}
Table~\ref{tab:participants} summarizes participant backgrounds. All nine designers were active LLM users ($M = 5.33/7$, $SD = 1.66$) with AI image generation experience ($M = 5.00/7$, $SD = 1.80$).

\begin{table}[!htbp]
\centering
\caption{Participant demographics.}
\label{tab:participants}
\setlength{\tabcolsep}{2pt}
\begin{tabular}{clllcc}
\toprule
\textbf{ID} & \textbf{Domain} & \textbf{Role} & \textbf{Gender} & \textbf{Age} & \textbf{Exp.} \\
\midrule
P1 & Graphic design & Designer & F & 27 & 5 \\
P2 & Interior / Furniture & Designer & F & 26 & 4 \\
P3 & Spatial design & Designer & M & 29 & 6 \\
P4 & Retail / Store planning & Planner & F & 26 & 5 \\
P5 & Exhibition \& Branding & Senior planner & F & 27 & 5 \\
P6 & Automotive / Product & Product planner & F & 26 & 4 \\
P7 & Spatial / Architecture & Designer & M & 26 & 4 \\
P8 & BX / Space planning & Planner & F & 26 & 5 \\
P9 & Architecture & Designer & M & 26 & 5 \\
\bottomrule
\end{tabular}
\end{table}

Figure~\ref{fig:survey} presents the full self-report survey results across all three conditions, including NASA-TLX, Intent Expression, Satisfaction, Output Quality, and Agency/Ownership items referenced throughout Section \ref{sec:6}.

\renewcommand{\thefigure}{B\arabic{figure}}
\setcounter{figure}{0}

\begin{figure*}[!htbp]
  \renewcommand{\thefigure}{B\arabic{figure}}
  \centering
  \includegraphics[width=\textwidth]{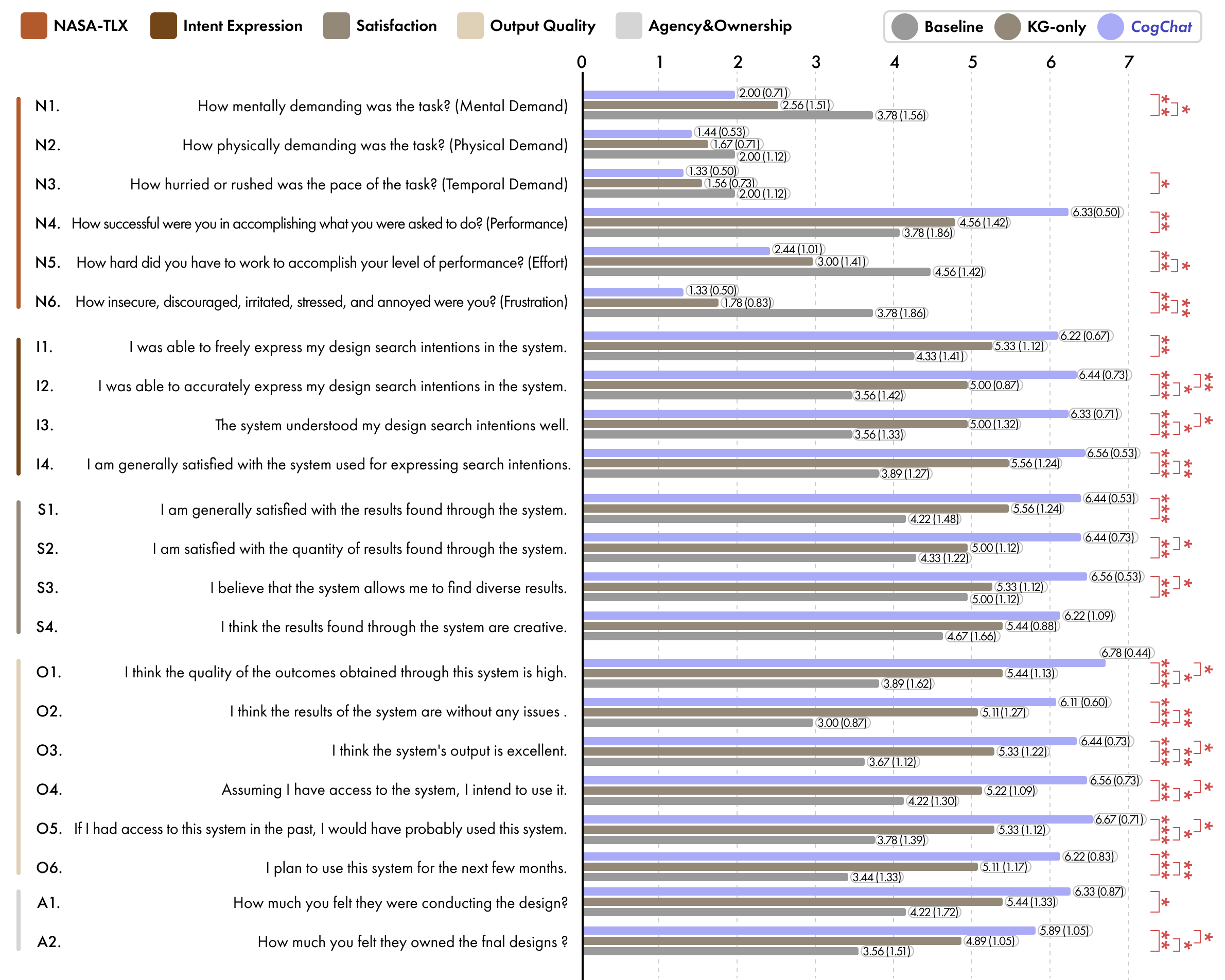}
  \caption{Self-report survey results across conditions. Horizontal bars show mean ($SD$) for NASA-TLX (N1--N6), Intent Expression (I1--I4), Satisfaction (S1--S4), Output Quality and Behavior Intention (O1--O6), and Agency and Ownership (A1--A2). Pairwise significance brackets: * $p<.05$, ** $p<.01$, *** $p<.001$ (Wilcoxon signed-rank, Benjamini--Hochberg corrected).}
  \Description{Horizontal bar chart of self-report survey results across three conditions. NASA-TLX workload subscales (N1--N6) show Baseline highest and CogChat lowest. Intent Expression (I1--I4), Satisfaction (S1--S4), Output Quality (O1--O6), and Agency and Ownership (A1--A2) show CogChat highest across all items. Pairwise significance brackets indicate Wilcoxon signed-rank tests with Benjamini-Hochberg correction.}
  \label{fig:survey}
\end{figure*}







\end{document}